\documentclass{emulateapj}
\usepackage{natbib}
\lefthead{Dewangan et al.}

\slugcomment{Submitted to ApJ Letters}

\newcommand\chandra{{\it Chandra}}
\newcommand\suzaku{{\it Suzaku}}

\newcommand\xmm{{\it XMM-Newton}}

\newcommand\ks{{\rm~ks}}
\newcommand\kev{{\rm~keV}}
\newcommand\ev{{\rm~eV}}
\newcommand\kms{\ifmmode {\rm~km\ s}^{-1} \else ~km s$^{-1}$\fi}
\newcommand\Hunit{\ifmmode {\rm~km\ s}^{-1}\ {\rm Mpc}^{-1}
        \else ~km s$^{-1}$ Mpc$^{-1}$\fi}
\newcommand\ctssec{\ifmmode {\rm~count\ s}^{-1} \else ~count s$^{-1}$\fi}
\newcommand\ergsec{\ifmmode {\rm~erg\ s}^{-1} \else
        ~erg s$^{-1}$\fi}
\newcommand\funit{\ifmmode {\rm~erg\ s}^{-1}\;{\rm cm}^{-2} \else
        ~ergs s$^{-1}$ cm$^{-2}$\fi}
\newcommand\phflux{\ifmmode {\rm~photon\ s}^{-1}\;{\rm cm}^{-2}
        \else   ~photon s$^{-1}$ cm$^{-2}$\fi}
\newcommand\efluxA{\ifmmode {\rm~erg\ s}^{-1}\;{\rm cm}^{-2}\;{\rm
        \AA}^{-1} \else ~erg s$^{-1}$ cm$^{-2}$ \AA$^{-1}$\fi}
\newcommand\efluxHz{\ifmmode {\rm~erg\ s}^{-1}\;{\rm cm}^{-2}\;{\rm
        Hz}^{-1} \else ~erg s$^{-1}$ cm$^{-2}$ Hz$^{-1}$\fi}
\newcommand\cc{\ifmmode {\rm~cm}^{-3} \else cm$^{-3}$\fi}
\newcommand\FWHM{\ifmmode {\rm~FWHM} \else ${\rm~FWHM}$\fi}
\newcommand\Msun{\ifmmode M_{\odot} \else $M_{\odot}$\fi}
\newcommand\Lsun{\ifmmode L_{\odot} \else $L_{\odot}$\fi}

\newcommand\gtsim{\raisebox{-.5ex}{$\;\stackrel{>}{\sim}\;$}}
\newcommand\hbeta{\ifmmode {\rm H}\beta \else H$\beta$\fi}
\newcommand\Kalpha{\ifmmode {\rm K}\alpha \else K$\alpha$\fi}
\newcommand\nh{\ifmmode N_{\rm H} \else N$_{\rm H}$\fi}
\usepackage{graphicx}

\begin{document}

\title{X-ray spectral cut-off and the lack of hard X-ray emission from
  two ultraluminous X-ray sources M81~X--6 and Holmberg~IX X--1}

\author{G. C. Dewangan\altaffilmark{1}, V. Jithesh\altaffilmark{2},
  R. Misra\altaffilmark{1} and C. D. Ravikumar\altaffilmark{2}} \altaffiltext{1}{Inter-University Centre
  for Astronomy and Astrophysics, Post Bag 4, Ganeshkhind,
  Pune-411007, India; gulabd@iucaa.ernet.in}

\altaffiltext{2}{Department of Physics, University of Calicut,
  Malappuram-673635, India}

\begin{abstract}

  We present broadband X-ray spectral study of two ultraluminous X-ray
  sources (ULXs), M81~X--6 and Holmberg~IX~X-1 based on \suzaku{} \&
  \xmm{} observations. We perform joint broadband spectral analysis of
  the brightest sources in the field, i.e. the two ULXs and the active
  galactic nucleus (AGN) in M81, and demonstrate that the X-ray
  spectra of the ULXs cut off at energies $\gtsim 3\kev$ with
  negligible contribution at high energies in the \suzaku{} HXD/PIN
  band. The $90\%$ upper limit on the $10-30\kev$ band luminosity of
  an underlying broadband power-law component is $3.5\times
  10^{38}{\rm~ergs~s^{-1}}$ for M81~X--6 and
  $1.2\times10^{39}{\rm~ergs~s^{-1}}$ for Holmberg~IX~X--1. These
  limits are more than an order of magnitude lower than the bolometric
  ($0.1-30\kev$) luminosity of $6.8\times10^{39}{\rm~ergs~s^{-1}}$ for
  M81~X--6 and $1.9\times 10^{40}{\rm~ergs~s^{-1}}$ for
  Holmberg~IX~X--1.  Our results confirm earlier indications of
  spectral cut-offs inferred from the \xmm{} observations of bright
  ULXs and show that there isn't an additional high energy power-law
  component contributing significantly to the X-ray emission. The
  spectral form of the two ULXs are very different from those of Galactic
  black hole X-ray binaries (BHBs) or  AGNs. This
  implies that the ULXs are neither simply scaled-up versions of stellar
  mass BHBs or scaled-down versions of AGNs.

\end{abstract}

\section{Introduction} 
Generally galaxies host numerous X-ray sources. Most off-nuclear,
compact X-ray sources are believed to be X-ray binaries (XRBs) with a compact
object -- white dwarf, neutron star or a black hole. A few of the
brightest X-ray sources at the high end of the point source luminosity
function can exceed the Eddington limit of even the most massive
stellar mass black hole, sometimes by large factors. The nature of
these sources, known as the ultra-luminous X-ray sources (ULXs),
continues to remain an enigma since their dynamical mass measurements have not
been possible.

Possible explanations for the high luminosities of ULXs include ($i$)
X-ray binaries with intermediate mass ($M_{BH} \simeq 10^2 -
10^4{\rm~M_{\odot}}$) black holes (IMBH)
\citep[e.g.,][]{1999ApJ...519...89C,2004IJMPD..13....1M}, ($ii$) X-ray
binaries  with anisotropic emission \citep{2001ApJ...552L.109K},
($iii$) beamed XRBs with relativistic jets directly pointing towards us
i. e., scaled down versions of blazars
\citep{1999ARA&A..37..409M,2002A&A...382L..13K}, and ($iv$) XRBs with
super-Eddington accretion rates
\citep{2001ApJ...551..897B,2002ApJ...568L..97B}.
\chandra{} and \xmm{} observations of ULXs have shown a variety of
spectral shapes e.g., simple powerlaw similar to the low/hard state of
BHBs \citep{2006ApJ...649..730W} and soft excess (blackbody $kT\sim
0.1-0.4\kev$) plus powerlaw component.  The soft X-ray excess emission
has been interpreted as optically thick emission from a thin
accretion disk with temperatures in the range of $\sim 100-300\ev$,
suggestive of an IMBH accreting at sub-Eddington ($\sim 0.1 L_{Edd}$)
rates
\citep{2003ApJ...585L..37M,2004ApJ...607..931M,2004ApJ...614L.117M,2004IJMPD..13....1M,2004MNRAS.355..359F}. However, the
cool disk plus powerlaw spectra of ULXs may not always correspond to
the high/soft state of BHBs \citep{2005MNRAS.357.1363R}. High S/N
\xmm{} observations of bright ULXs has revealed curvature or cut-offs
in the spectra at high energies $\gtsim 5\kev$
\citep{2006MNRAS.368..397S,2006ApJ...641L.125D,2006ApJ...638L..83A,2009MNRAS.397..124G}. However,
it is not known if the X-ray spectrum actually has a cut-off or there
is a deficit of emission near $7\kev$ due to perhaps an absorption
edge.  Observation of strong X-ray emission above $10\kev$ will put
strong constraints on the nature of ULXs. The presence of a strong
hard X-ray powerlaw will establish ULXs to be either scaled-up BHBs or
scaled-down versions of AGNs and the lack of emission near $7\kev$ may
then be explained as the blurred iron K-edge due to Compton reflection
arising from regions very close to a black hole. On the other hand,
absence of such a strong hard X-ray component will make ULXs very
different from BHBs or AGNs, probably accreting in a state that is not
usually attained by them.

\suzaku{} observations have shown that the broadband spectrum of
M82~X--1 is slightly curved \citep{2009PASJ...61S.263M}.  However,
M~82~X--1 is in a crowded field and it is not clear if the
contribution of nearby multiple unresolved sources both in the XIS and
PIN FOV can alter the broadband spectral shape.  Here we study the
broadband spectral shape of two bright ULXs M81 X--6 and Holmberg IX
X--1 (Ho~IX~X--1) located in the dwarf
irregular galaxy Holmberg IX, a companion of M81. 
Ho~IX~X--1 is also known as M81~X--9. We constrain the
$0.6-30\kev$ spectral shape of the two ULXs for the first time based
on joint spectral modeling of multiple sources observed with
\suzaku{}.  

\section{Observations and Data Reduction}

We used the onaxis \suzaku{} observations of M81 (observation ID:
906004010) and Holmberg IX X--1 (observation ID: 707019010) performed
during 15--16 September 2011 and 13--17 April 2012 for exposure times
of $45.6\ks$ and $182.5\ks$, respectively. We also used \xmm{}
observation of Holmberg IX X--1 performed during 26-27 September 2004
for an exposure time of $119.1\ks$ (observation ID : 0200980101).

\begin{figure*}
  \centering
  \includegraphics[width=5.5cm]{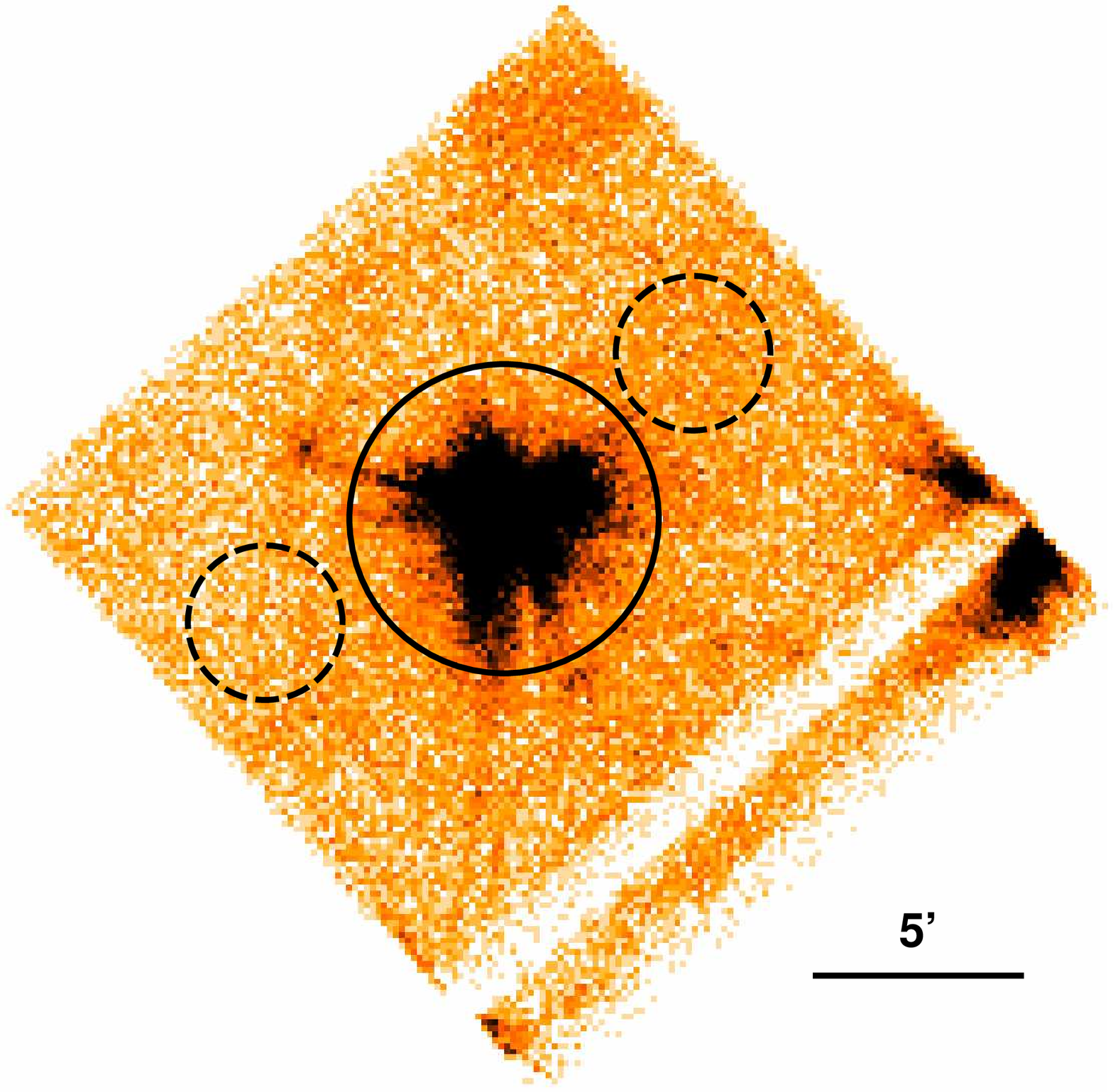}
  \includegraphics[width=5.5cm]{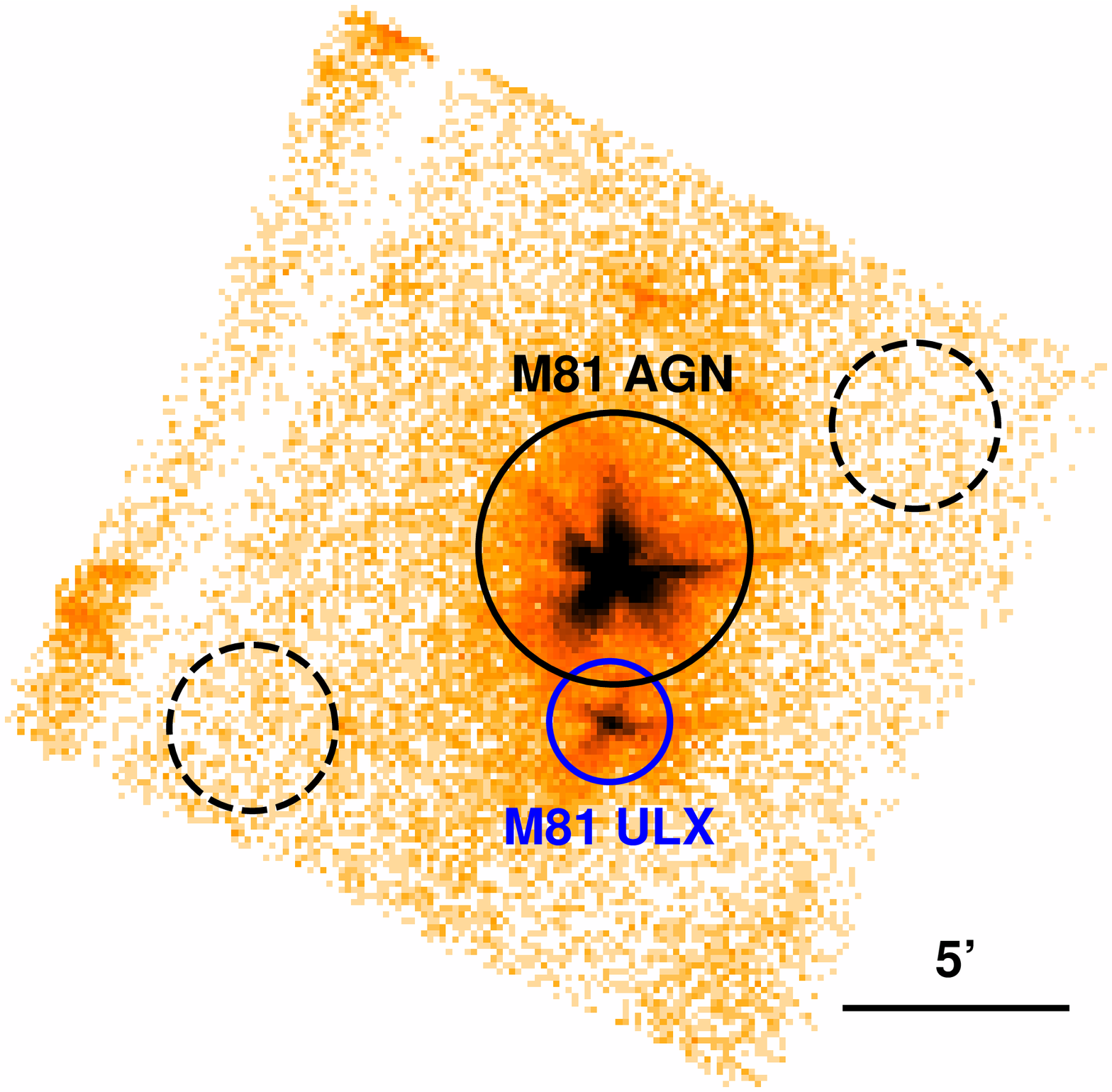}
  \includegraphics[width=5.8cm]{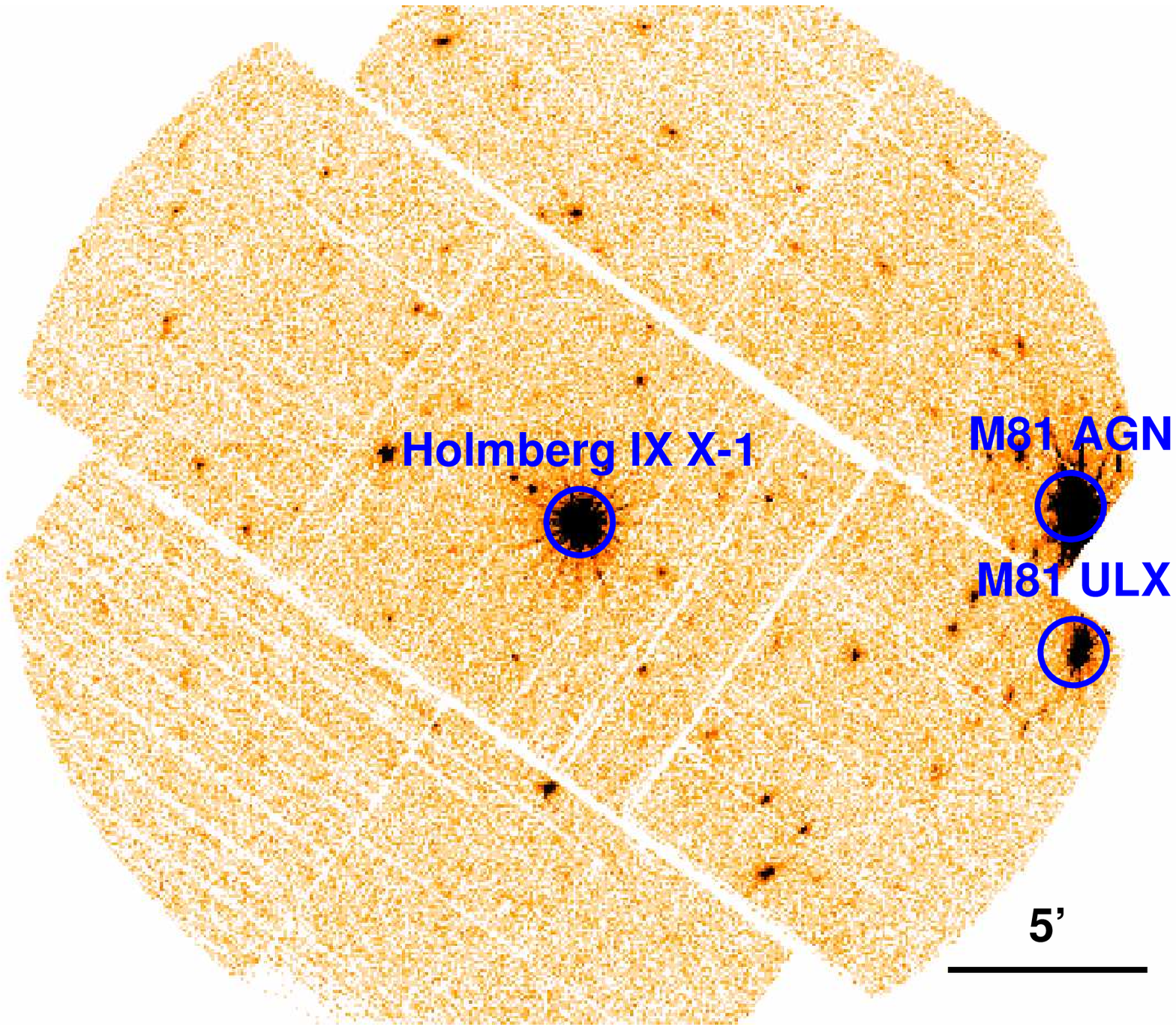}
  \caption{\suzaku{} XIS images of Holmberg IX X-1 ({\it left panel})
    and M81 AGN and ULX ({\it middle panel}) and the \xmm{} MOS image
    of the region containing Ho IX X--1, the ULX and AGN in M81. Also
    shown are the circular regions with solid lines for the sources
    and dashed line for background that were used for spectral
    extraction. }
  \label{ulx_images}
\end{figure*}

We processed the unfiltered \suzaku{} data with {\tt aepipeline}
available with {\tt HEASOFT version 6.13} and used the recent
calibration database (CALDB) available as of 20 January 2013.  We
extracted the spectra of Ho IX X--1 from the filtered XIS event lists
using a circular region of radius $220\arcsec$, and the corresponding
background spectra from two circular regions each with radius
$110\arcsec$.  The ULX M81~X--6 is $3.4\arcmin$ away from the
AGN. Therefore, we extracted the XIS spectra for M81 AGN and M81 X--6
using smaller circular regions of radii $180\arcsec$ and $80\arcsec$,
respectively. We extracted common background spectra applicable to
both AGN and ULX using two circular regions of radius $110\arcsec$
each. The circular extraction regions used to extract the spectral
products are shown in Figure~\ref{ulx_images}. The XIS0 and XIS3 are
the front illuminated (FI) CCDs and the two spectra for each object
were combined using the tool {\it addascaspec}.  We also extracted the
HXD/PIN source and background spectra from both the observations using
the task {\it hxdpinxbpi} from the PIN event lists and the tuned non
X-ray background files appropriate for the observations.

Since the HXD/PIN is a non-imaging instrument with a field of view
(FOV) of $34\arcmin\times 34\arcmin$ FWHM, there may be contribution
from sources within the FOV. To identify the potential contaminating
sources, we analyzed the \xmm{} observation of Ho~IX~X01. We processed
the EPIC-pn and MOS data in a standard manner. We show the MOS image
of the field containing Ho IX X-1 in Figure~\ref{ulx_images}.  In the
MOS FOV of $30\arcmin$ diameter, apart from Ho~IX~X-1., there are two bright sources, M81 AGN and  ULX M81 X--6 which  can contribute to the PIN spectrum.

\begin{table}
  \caption{Best-fit parameters for the ULXs and AGN.}
  \begin{tabular}{llccc} \hline
    Comp. & Par. & AGN+M81ULX & \multicolumn{2}{c}{AGN+Ho~IX~X--1} \\ \hline
    &          & Model 1$~^a$        & Model 2$~^a$ & Model 3$~^a$ \\ \hline
    Abs.       & $N_H^{Gal}~^{(b)}$ & $5.5$(fixed)  &  $5.5$(fixed) & $5.5$(fixed) \\
    PL         & $\Gamma$  & $1.91_{-0.02}^{+0.02}$ &  $1.91_{-0.03}^{+0.02}$   & $1.89_{-0.02}^{+0.03}$ \\
    & $n_{PL}~^{(c)}$  & $4.4_{-0.1}^{+0.2}$ &  $4.4_{-0.1}^{+0.2}$ & $4.3_{-0.1}^{+0.1}$ \\
    GL         & $E_{line}$(keV)    & $6.57_{-0.07}^{+0.09}$ &  $6.57_{-0.08}^{+0.09}$  & $6.57_{-0.09}^{+0.08}$  \\
    & $\sigma$(eV)  & $176_{-75}^{+91}$ &  $175_{-75}^{+91}$ &  $162_{-75}^{+90}$ \\
    & $f_{line}~^{(d)}$    & $1.4_{-0.4}^{+0.6}$   & $1.4_{-0.4}^{+0.6}$ & $1.3_{-0.5}^{+0.5}$\\
    bbody            & $kT_{bb}$(eV) &  -- &  $260_{-13}^{+14}$ & --      \\ 
    diskbb        & $kT_{in}$(eV) &  -- &  --  & $353_{-60}^{+49}$ \\
    Cutoff PL        & $\Gamma$ & $0.6_{-0.1}^{+0.2}$ & $1.1_{-0.1}^{+0.1}$ & -- \\
    & $E_{cutoff}$(keV)  & $2.7_{-0.3}^{+0.4}$  & $8.0_{-1.3}^{+1.9}$ & -- \\
    & $n_{cpl}~^{(e)}$  & $7.2\pm0.3$   & $11.5_{-0.6}^{+0.9}$ & --\\
    nthcomp    &   $\Gamma$ &  -- & -- &  $1.64_{-0.03}^{+0.02}$  \\
    & $kT_e$(keV) & -- & -- &  $2.6_{-0.2}^{+0.2}$ \\
    & $kT_S$(eV)  & -- &  -- &  $kT_{in}$ \\
    GL         & $E_{line}$(keV)  & -- &    $6.32_{-0.58}^{+0.15}$ & $6.32_{-0.27}^{+0.18}$ \\
    & $\sigma$(eV) & -- &  $227$(fixed) &  $<495$ \\
    & $f_{line}~^{(d)}$  & -- &   $0.04\pm0.02$  & $0.026_{-0.018}^{+0.022}$ \\ 
    & $\chi^2/dof$ &  418.4/435 &  509/517   & 519/517   
    \\ \hline
  \end{tabular}

  $^{(a)}$ Model 1: AGN ({\tt wabs$\times$(PL+GL(1))}) + ULX ({\tt wabs$\times$cutoffpl}); Model 2: AGN ({\tt wabs$\times$(PL+GL(1))}) + ULX ({\tt wabs$\times$(bbody + cutoffpl + GL(2)}); Model 3: AGN ({\tt wabs$\times$(PL+GL(1))}) + ULX ({\tt wabs$\times$(diskbb + nthcomp + GL(2)}). $^{(b)}$ Galactic column in units of $10^{-20}{\rm~cm^{-2}}$. 
  $^{(c)}$ Normalization in   $10^{-3}{\rm~photons~cm^{-2}~s^{-1}~keV^{-1}}$ at $1\kev$. 
  $^{(d)}$ Line flux in  $10^{-4}{\rm~photons~cm^{-2}~s^{-1}}$. 
  $^{(e)}$ Normalization of  cutoff PL
  in   $10^{-4}{\rm~photons~cm^{-2}~s^{-1}~keV^{-1}}$ at $1\kev$.
  \label{tab1}
\end{table}

\section{Analysis \& Results}
We begin with the spectral analysis of the \suzaku{} observation of
M~81. While the AGN and the ULX are clearly resolved in the XIS data,
the PIN data is additionally contaminated by the nearby bright ULX
Ho~IX~X--1. Other sources seen in the \xmm{} MOS image are too faint
to contribute to the PIN spectrum. We fitted an absorbed powerlaw
model to the PIN and the AGN XIS data. We used the the $16-30\kev$ band 
PIN data (as there is no detection above $30\kev$) and the $2-10\kev$ XIS band
data which excludes the low energy thermal emission of the AGN
\citep{2003A&A...400..145P}. M~81 AGN is known to show an iron line
\citep{2004ApJ...607..788D} and therefore we also used a Gaussian
line.  This simple model resulted in a good fit, $\chi^2 = 245$ for
$290$ degrees of freedom (dof). To find any contribution from the
nearby bright ULXs to the HXD/PIN, we included an additional power-law
component. We fixed the normalization of the second powerlaw component
to zero for the XIS and varied it for the PIN data. This did not improve
the fit and hence we calculated $90\%$ upper-limit on the $10-30\kev$ flux
by fixing the photon index $\Gamma_2$ at different values ranging from
1 to 2.5. We find that the maximum possible contribution by the nearby ULXs is
$<6.8\times10^{-13}{\rm~ergs~cm^{-2}~s^{-1}}$ in the $10-30\kev$ band.

To estimate the hard X-ray contribution from the ULXs, 
we fitted the XIS spectra
of Ho~IX X--1 and M81 X--6 with absorbed power-law models and
calculated the $10-30\kev$ band fluxes by extrapolating the best-fit
models. We found
$f_X$($10-30\kev$)$=8.3_{-0.2}^{+0.1}\times10^{-12}{\rm~ergs~cm^{-2}~s^{-1}}$
for Ho~IX~X--1 and $(3.6\pm0.2)\times10^{-12}$ for M81 X--6. Thus, the
estimated fluxes for the ULXs based on the extension of the 
simple power-law models are
much higher than the $90\%$ upper-limit on the contribution by the
nearby sources to the HXD/PIN data. This implies that the ULXs lack significant
hard X-ray emission and their spectra cannot be described by a single
power-law from soft-to-hard X-rays. This confirms the earlier results of
a high energy cutoff in the \xmm{} spectra of these ULXs
\citep{2006ApJ...641L.125D,2006MNRAS.368..397S}.

To better constrain the broadband X-ray spectral shapes of the
ULXs, we performed joint spectral analysis of the \suzaku{} XIS+PIN
spectra of the ULXs and the AGN. For this purpose, we created ISIS fit
functions which are a combination of models for the ULXs and the AGN
but a given model component is applied to a specific spectrum or a
subset of spectra. First we fit the spectra obtained from the onaxis
\suzaku{} observation of the M81 AGN. We used the combined XIS FI
spectra of the M81 ULX and the AGN and the corresponding HXD/PIN
spectrum of the field. We did not use XIS spectrum of Ho IX X--1 as it
is not in the field of view. We also did not account for its
contribution to the PN spectrum. This ULX is $12.4\arcmin$ away from
the AGN in M81. Thus, given the triangular response of HXD/PIN with an
FWHM of $34\arcmin$, the ULX contributes $\sim 50\%$ of its actual
flux to the PIN data. As inferred above, Ho IX X--1 generally lacks
the hard X-ray emission and its contribution to the PIN band is likely
to be very small. We further deal with this issue below when we
analyze the on-axis \suzaku{} observation of Ho~IX~X--1.

\begin{figure*}
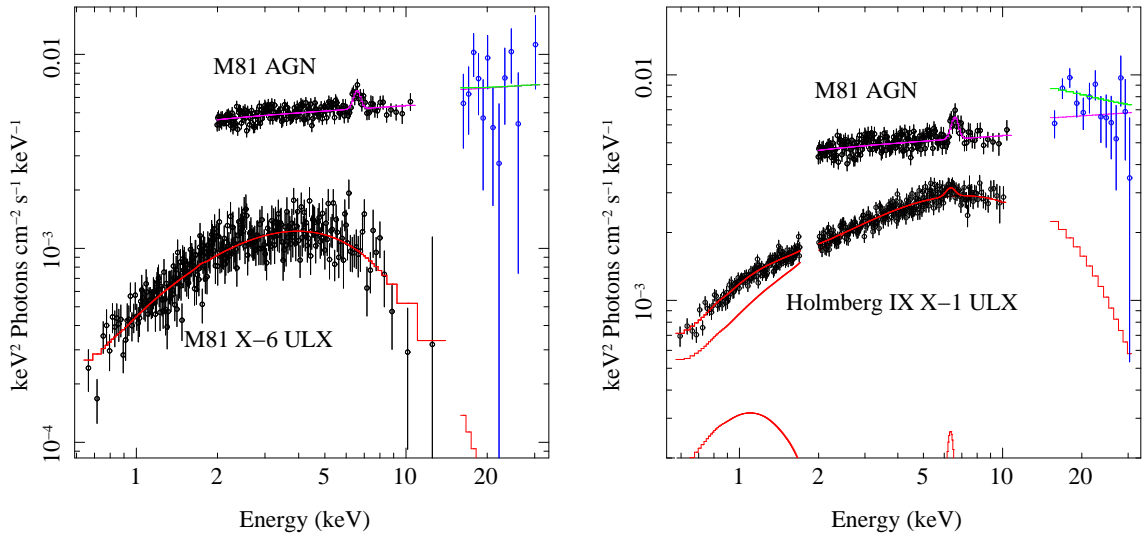

  \centering
  \includegraphics[width=7cm,angle=-90]{fig2a.ps}
  \includegraphics[width=7cm,angle=-90]{fig2b.ps}
  \caption{Unfolded XIS and PIN spectra of ULXs and AGN. {\it Left
      panel:} The XIS and PIN spectral data and the best-fit models
    for the AGN and the ULX in M81 derived from the onaxis \suzaku{}
    observation of M81 on 15 September 2011. {\it Right panel:} The
    XIS spectrum of M81 AGN obtained from the \suzaku{} observation of
    15 September 2011, the XIS spectrum of Ho IX X-1 and the
    corresponding PIN spectrum derived from the onaxis \suzaku{}
    observation of Ho IX X--1 on 13 April 2012. Also shown are the
    individual spectral component -- blackbody, Gaussian line and the
    cutoffpl for Ho~IX~X--1.}
  \label{jointfit}
\end{figure*}

For the joint spectral analysis of the M81 ULX and the AGN spectra, we
used a joint model which is a combination of the model for the ULX
({\tt wabs$\times$(cutoffpl + gauss)}) and the AGN ({\tt
  wabs$\times$(powerlaw + gauss)}).  We refer to this as model 1. The ULX
model only is applied to XIS spectrum of M81 X--6 and the AGN model
alone applied to the XIS spectrum of the AGN but both 
models are combined to fit to the PIN data. 
As before, we excluded XIS data below
$2\kev$ for the AGN only and used a constant multiplicative factor of
$1.16$ for the PIN data. The model provided a statistically acceptable
fit with $\chi^2=418.4/435$ and the best-fit parameters are listed in
in Table~\ref{tab1} while the unfolded spectral data and the best-fit
model are shown in Figure~\ref{jointfit}. 
We also examined the
continuation of the powerlaw in the hard band by fixing $E_{cutoff}$
at $100\kev$. The fit became worse with $\chi^2/dof =679/436$, ruling
out the presence of such a powerlaw. Further, we added an additional
powerlaw component with $\Gamma=2$ to the best-fit ULX model 1 and
estimated the $90\%$ upper-limit on the $10-30\kev$ flux to be $f_X =
2.3\times10^{-13}{\rm~ergs~cm^{-2}~s^{-1}}$ and the corresponding
luminosity $L_X = 3.5\times10^{38}{\rm~ergs~s^{-1}}$ for an underlying
power-law component.
We also tested physical models that do not contribute significantly in
the hard X-ray band.  We replaced the {\tt cutoffpl} in model 1 with
the standard disk model {\tt diskbb} which resulted in a good fit
($\chi^2/dof = 432.4/436$) with $kT_{in} = 1.62_{-0.05}^{+0.04}\kev$,
$f_X(0.1-10\kev)=(3.7\pm0.1)\times10^{-12}{\rm~ergs~cm^{-2}~s^{-1}}$.
We also tested the p-free disk model {\tt diskpbb} which allows the
disk temperature to scale with radius as $r^{-p}$. For standard disk
model, $p=0.75$ and lower values of $p$ may indicate a slim accretion disk
\citep{2000PASJ...52..133W}. The
p-free disk model resulted in a slightly better fit ($\chi^2/dof =
421.5/435$) with $kT_{in} =1.8\pm0.1$, $p=0.69\pm0.02$ and
$f_X(0.1-30\kev) = (4.3\pm0.1)\times10^{-12}{\rm~ergs~cm^{-2}~s^{-1}}$
and $L_X((0.1-30\kev)=6.8\times10^{39}{\rm~ergs~s^{-1}}$. The
parameters of the AGN model components remained similar for different
ULX models {\tt cutoffpl}, {\tt diskbb} and the p-free disk.  
Thus, we confirm that the M81 X--6 spectrum cuts-off at high energies
and does not contribute significantly to the HXD/PIN band. Moreover,
any possible hard X-ray contribution by the ULX Ho IX X--1
(which we show below is unlikely) would further strengthen the result that
 M81 X--6 is dim in hard X-rays.

We next perform the joint spectral analysis of the onaxis \suzaku{}
observation of Ho~IX~X--1. The ULX is well isolated spatially and its
X-ray emission in the XIS band is not contaminated by any other strong
source (see Fig.~\ref{ulx_images}). However, the PIN data is obviously
contaminated by the AGN in M81. We used a joint spectral model similar
to the model 1 that combines the ULX model ({\tt wabs$\times$(bbody +
  cutoffpl + gauss}) and the AGN model ({\tt wabs$\times$ (powerlaw +
  gauss)}). We refer this as model 2.  As before we used a relative
normalization of $1.16$ between the XIS spectrum of Ho~IX~X--1 and its
contribution to the PIN spectrum. Since there is no simultaneous XIS 
spectrum of the AGN, we used the AGN XIS
spectrum from the onaxis observation of M81. We used a
variable relative normalization between the XIS spectrum of the AGN
and its contribution to the PIN band to allow for flux variability of
the AGN between the two \suzaku{} observation. Thus, the use of the
XIS AGN spectrum in modeling the ULX spectrum is equivalent to
assuming that the AGN did not change its spectral shape. This is 
justified as the
$2-10\kev$ spectral shape of M81 AGN is known to show very little
variability if at all \citep{1996PASJ...48..237I}. As before we
excluded the XIS spectrum of AGN below $2\kev$. The joint fit resulted
in a statistically acceptable fit ($\chi^2/dof = 509.3/517$). 
The best-fit parameters are listed in Table~\ref{tab1} while  The unfolded
spectrum and the best-fit models for the ULX and AGN are shown in
Figure~\ref{jointfit}.
The best-fit parameters for the AGN were
similar to that obtained earlier (see Table~\ref{tab1}). 
 As expected from the earlier analysis, the
spectrum of Ho~IX~X--1 cuts-off and makes very little contribution to
the HXD/PIN band. A weak iron line is detected for $\Delta\chi^2 =
-11.5$ for two additional parameters i.e., at a significance level of
$99.7\%$ according to the F-test. The line width was not well
constrained, therefore $\sigma$ was fixed at its best-fit value. Such
a weak iron line was not detected in the long \xmm{} observation
\citep{2006ApJ...641L.125D} and is most likely associated with the
host galaxy.  The relative normalization for the AGN contribution to
the PIN data is $0.58_{-0.38}^{+0.41}$. Fixing the value to $1.16$ does not
change the cutoff energy for the ULX and hence uncertainties in the AGN flux
does not effect the overall result.
We further tested the reality of the spectral cut-off of
Ho~IX~X--1 by fixing the cutoff energy at $100\kev$. This resulted in
a poor fit ($\Delta \chi^2 = +99.4$ for the loss of only one
parameter). Thus, the presence of spectral cut-off is highly
significant. As before, we estimated $90\%$ upper limit on the
$10-30\kev$ flux of an underlying powerlaw component to be $f_X =
7.5\times10^{-13}{\rm~ergs~cm^{-2}~s^{-1}}$ and the corresponding
luminosity to be $L_X = 1.2\times10^{39}{\rm~ergs~s^{-1}}$. We also
tested the thermal Comptonization model by replacing the {\tt
  cutoffpl} in model 2 with the {\tt nthcomp} and {\tt bbody} with the
{\tt diskbb} component (model 3), the joint fit resulted in
$\chi^2/dof = 519/517$) with $kT_e = 2.6\pm0.2\kev$. In this fit, we
treated the {\tt diskbb} component with $kT_{in} \sim 0.35\kev$ to be
the seed photon source for the Comptonization. We calculated the
flux of the {\tt diskbb} and {\tt nthcomp} components to be
$5.4\times10^{-13}{\rm~ergs~cm^{-2}~s^{-1}}$ and
$1.2\times10^{-11}{\rm~ergs~cm^{-2}~s^{-1}}$, respectively.

\section{Discussion}
Using the broadband \suzaku{} observations of the fields containing
two bright ULXs, we measured X-ray luminosities of
$6.8\times10^{39}{\rm~ergs~s^{-1}}$ for M81 X--6 and
$1.9\times10^{40}$ for Ho~IX~X--1. With the joint spectral analysis of
XIS and PIN spectra of the brightest sources in the field, we showed
that the broadband spectra of M81~X--6 and Ho IX X--1 cutoff at
$2.8\kev$ and $8\kev$, respectively. This conforms the high energy
spectral curvature in these ULXs earlier found with the limited
bandpass of \xmm{} \citep{2006ApJ...641L.125D,2006MNRAS.368..397S}. In
addition, we do not find additional high energy spectral components
similar to that observed from BHBs and AGNs. This demonstrates clearly
that the two ULXs are in unusual spectral states that are distinctly
different than that observed from BHBs and AGNs. 

The spectrum of M81~X--6 is well described by the standard disk model
with high temperature $kT_{in} = 1.62_{-0.05}^{+0.04}\kev$ similar to
that observed from BHBs but the inferred X-ray luminosity exceeds the
Eddington luminosity of $44M_{\odot}$ black hole. Generally BHBs show
standard disk spectra in their high/soft state with the relative
accretion rate in the range of $\sim 0.1-0.5$
\citep{1997ApJ...489..865E}.  Thus, the observed X-ray luminosity of
M81 X--6 will correspond to $\gtsim 100M_{\odot}$. The temperature of
the standard disk around such a large black hole is unlikely to exceed
$1\kev$. Thus, M81 X--6 is not in the high/soft state generally
observed in BHBs. The p-free disk fit showed that the temperature of
disk ($T \propto r^{-0.68}$) at different radii is lower than that
expected from a standard disk ($T \propto r^{-0.75}$). This suggests
that the ULX M81 X--6 likely hosts a slim accretion disk.

We confirmed the soft excess from Ho~IX~X--1 and showed that the
spectral downturn near $8\kev$ earlier observed with \xmm{} continues
in the hard X-ray band above $10\kev$.  The soft excess and the
spectral turnover is well described by a standard cool disk and cool
Comptonization as found earlier
\citep{2006ApJ...641L.125D,2006MNRAS.368..397S}. Based on the spectral
analysis of high quality X-ray spectra of ULXs observed with \xmm{},
\citet{2009MNRAS.397..124G} showed that most ULX spectra are best
described by an outer cool disk and optically thick, cool
Comptonization. They termed such a spectral form as the ultraluminous
state which is thought to represent spectral transition between
Eddington and super-Eddington accretion flows. It is interesting to
note that the luminosity of the soft excess component, described as
{\tt diskbb}, is $8.4\times10^{38}{\rm~ergs~s^{-1}}$ corresponding to
relative accretion of $\dot{m}/{\dot{m_{edd}}} \sim 0.2$ for a
$30M_{\odot}$ black hole. Thus, if there is mild beaming of the
Comptonized component, a stellar mass black hole can explain the
observed luminosity of Ho~IX X--1.

Finally, we note that while many BHBs and AGNs show strong
relativistically broadened iron line, we are yet to observe such lines
from any ULX. \citet{2010MNRAS.402.2559C} explained the soft excess
and drop near $7\kev$ in the spectra of several ULX including M81 X--6
and Ho IX X--1 in terms of relativistically blurred Compton reflection
from accretion disks. Such a model requires strong continuum above
$7.1\kev$. However, our finding of the spectral downturn and its
continuation in the hard band above $10\kev$ in Ho X--1 and M81 X--6
imply the absence of a strong illuminating continuum.  Thus, our
findings rule out the possibility of explaining the spectral downturn
near $7\kev$ as a blurred iron K-edge. The absence of a strong
illuminating continuum is most likely the reason behind the
non-detection of relativistically broadened iron lines from ULXs.

\end{document}